\documentclass[12pt]{article}
\usepackage{epsf}
\def\be{\begin{equation}}
\def\ee{\end{equation}}
\def\bea{\begin{eqnarray}}
\def\eea{\end{eqnarray}}

\begin{document}
\begin{titlepage}
\begin{center}
{\Large \bf William I. Fine Theoretical Physics Institute \\
University of Minnesota \\}
\end{center}
\vspace{0.2in}
\begin{flushright}
FTPI-MINN-17/01 \\
UMN-TH-3617/17 \\
January 2017 \\
\end{flushright}
\vspace{0.3in}
\begin{center}
{\Large \bf Some properties of the states at the hidden-strangeness mixed ${1 \over 2}^+ + {1 \over 2}^-$ heavy meson pair threshold in $e^+e^-$ annihilation.
\\}
\vspace{0.2in}
{\bf  M.B. Voloshin  \\ }
William I. Fine Theoretical Physics Institute, University of
Minnesota,\\ Minneapolis, MN 55455, USA \\
School of Physics and Astronomy, University of Minnesota, Minneapolis, MN 55455, USA \\ and \\
Institute of Theoretical and Experimental Physics, Moscow, 117218, Russia
\\[0.2in]

\end{center}

\vspace{0.2in}

\begin{abstract}
The threshold behavior of $e^+ e^-$ annihilation is considered in the channels $B_{s0} \bar B_s^* + c.c.$, $B_{s1} \bar B_s + c.c.$, and $B_{s1} \bar B_s^* +c.c.$, where $B_{s0}$ and $B_{s1}$ are the excited bottom-strange $J^P=0^+$ and $J^P = 1^+$ mesons. It is argued that due to the heavy quark spin symmetry (HQSS) only one coherent combination of the first two channels is produced in the $S$ wave as well as the third channel. Thus, if there exist threshold molecular peaks in the considered channels, only two of such peaks can be formed in the annihilation. The properties of such threshold states are discussed, including the heavy-light spin structure and the related transitions to bottomonium plus light mesons, as well as mixing with the channels with and without hidden strangeness.

\end{abstract}
\end{titlepage}

The structures with hidden heavy flavor produced in the $e^+ e^-$ annihilation near the thresholds for heavy meson pairs are of a well known great interest for the studies of those mesons [e.g. $D$ mesons at $\psi(3770)$, or $B$ mesons at $\Upsilon(4S)$] as well as for exploring the strong interaction between the heavy hadrons that results in an intricate sequence of peaks and dips in the yield of specific final channels. The bulk of the data available thus far mostly relates to the production of pairs of two ground state mesons, i.e. the pairs $D^{(*)} \bar D^{(*)}$ in the charm sector, or $B^{(*)} \bar B^{(*)}$ in the beauty sector, with a possible addition of a pion, and significantly less is known about the excited heavy mesons, especially in the beauty sector. In particular, the strange excited $B_{s0}$ and $B_{s1}$ mesons with quantum numbers $J^P=0^+$ and $J^P=1^+$ are not yet observed, and their expected properties are predicted~\cite{beh,kl,gscpz} by extrapolation from those of their charm counterparts $D_{s0}(2317)$ and $D_{s1}(2460)$. It has been recently argued~\cite{bmv} that the strange excited $B$ mesons can be studied in the $e^+e^-$ annihilation in the Belle II experiment at energies near the thresholds for the channels $B_{s0} \bar B_s^* + c.c.$, $B_{s1} \bar B_s + c.c.$, and $B_{s1} \bar B_s^* +c.c.$, which thresholds are expected to be at approximately 11.1\,GeV for the first two final states and about 48\,MeV higher for the third one. The purpose of the present paper is to elaborate in some detail on the properties of the $J^{PC}=1^{--}$ states of this type produced in the $e^+e^-$ annihilation. In particular, it will be argued, based on HQSS, that only one superposition of the $C$-odd states $B_{s0} \bar B_s^* - c.c.$, $B_{s1} \bar B_s - c.c.$ is being produced, and that the heavy-light spin structure of this superposition and of the higher state $B_{s1} \bar B_s^* - c.c.$ is, to an extent, analogous to that in the $Z_b(10610$ and $Z_b(10650)$ resonances~\cite{bgmmv} (albeit in the sector with opposite parity), resulting in a comparable rate of hadronic transitions to levels of ortho- and para-bottomonium. Furthermore, it will be argued that the rescattering of the considered states to pairs of the ground-state mesons, both with and without hidden strangeness, is vanishing in the heavy quark limit, so that the yield of the $B_{s0}$ and $B_{s1}$ mesons should be more prominent than that for their charmed counterparts in the charmonium-like sector at 4.4 - 4.55\,GeV. 
In what follows is presented a brief overview of the orbitally excited heavy mesons and their possible production in the $e^+e^-$ annihilation with a subsequent discussion of the heavy-light spin structure of the produced states and of rescattering between the channels. Specific numerical estimates, based on the heavy quark expansion and the data on the charmed mesons, as well as recently acquired (and still scarce) data on excited non-strange $B$ mesons, is postponed till later in the paper, where also discussed are possible (dis)similarities with the observed behavior in the charmonium-like energy region.

As is well known, the ground-state heavy-light mesons form a doublet consisting of a pseudoscalar and a vector meson, e.g. $B$ and $B^*$, with the mass splitting inversely proportional to the heavy qaurk mass $M_Q$. The light antiquark in these mesons is an the $S_{1/2}$ state, corresponding to $J^P=(1/2)^-$. In the orbitally excited mesons the light antiquark can be in either $P_{1/2}$ ($J^P=(1/2)^+$) state or $P_{3/2}$ ($J^P=(3/2)^+$), which are split in energy by the spin-orbit interaction of the light quark, so that this splitting is generally of order $\Lambda_{QCD}$ and goes to a constant in the limit $M_Q \to \infty$. Accordingly, the orbitally excited heavy-light mesons come in two doublets: scalar $B_0$ and axial $B_1$, corresponding to the $(1/2)^+$ state of the light antiquark, and $B^*_1$ ($J^P=1^+$) and  $B^*_2$ ($J^P=2^+)$ mesons where the light antiquark is in the $(3/2)^+$ state. Generally, there can be a mixing between the axial $B_1$ and $B^*_1$ mesons, that is suppressed by HQSS, so that the separation between $(1/2)^+$ and $(3/2)^+$ axial mesons is expected to be much better in the beauty sector than for the charmed mesons. The non-strange excited $(1/2)^+$ mesons can decay in $S$ wave to a ground-state meson and a pion, e.g. $B_0 \to B \pi$, and are very broad ($\sim 300\,$MeV), while similar decays of the mesons in the $(3/2)^+$ doublet, e.g. $B_1 \to B^* \pi$, are required by HQSS to proceed in the $D$ wave, so that the latter mesons are relatively narrow. This property is quite different for the strange excited mesons, as is known for the $(1/2)^+$ $D_{s0}(2317)$ and $D_{s1}(2460)$, which are very narrow. The isospin-conserving decays into Kaon and a ground non-strange meson are kinematically impossible, so that their dominant decay~\cite{pdg} is an isospin-breaking one into a strange ground state $(1/2)^-$ meson and $\pi^0$ with a rate comparable to that of similar radiative transitions.

The heavy meson-antimeson pairs with one excited meson and one ground-state, are allowed, by quantum numbers, to be produced in the $S$ wave in $e^+e^-$ annihilation, thus suggesting a strong production near threshold. However, such production is forbidden for the combination $(3/2)^+ + (1/2)^-$ by HQSS~\cite{lv} and may proceed only due to symmetry breaking~\footnote{An $S$ wave production of $D^*_1 \bar D + c.c.$ was discussed in connection with the $Y(4260)$ charmonium-like resonance~\cite{cwghmz} and a similar production of $B^*_1 \bar B + c.c.$ was suggested~\cite{bv} at the peak $\Upsilon(6S)$.}. This behavior can be established by considering the heavy-light spin structure of the pairs of heavy mesons. Using the notation, for concreteness, for the beauty mesons, one can write the quark spin combinations in the mesons as
\bea
&& B \sim (b^\dagger q)~,~~ B_i \sim (b^\dagger \sigma_i q)~;~~ \bar B \sim (q^\dagger b)~,~~ \bar B_i \sim -(q^\dagger \sigma_i b) 
\label{dec} \\
&& B_0 \sim \left (b^\dagger (\sigma \cdot p) q \right )~,~~B_{1i} \sim \left ( b^\dagger \sigma_i (\sigma \cdot p) q \right ) ~;~~ \bar B_0 \sim \left ( q^\dagger (\sigma \cdot \bar p)  b \right ) ~,~~ \bar B_{1i} \sim - \left ( q^\dagger (\sigma \cdot \bar p) \sigma_i b \right ) \nonumber \\
&& B^*_{1i} \sim \sqrt{2} \, \left ( b^\dagger (p_ i + {i \over 2} \, \epsilon_{ijk} p_j \sigma_k) q \right )~,~~B^*_{2ij} \sim {1 \over 2 \, \sqrt{3}} \, \left ( b^\dagger  (3 \, p_i \sigma_j + 3 \, p_j \sigma_i - 2 \, \delta_{ij} (\sigma \cdot p) ) q \right )  \nonumber \\
&& {\bar B}^*_{1i} \sim \sqrt{2} \, \left ( q^\dagger (\bar p_ i + {i \over 2} \, \epsilon_{ijk} \bar p_j \sigma_k) b \right )~,~~\bar B^*_{2ij} \sim  {-1 \over 2 \, \sqrt{3}} \, \left ( q^\dagger  (3 \, \bar p_i \sigma_j + 3 \, \bar p_j \sigma_i - 2 \, \delta_{ij} (\sigma \cdot \bar p) ) b \right )\, , \nonumber 
\eea
where $\sigma_i$ are the Pauli matrices, $p_i$ ($\bar p_i$) is the momentum of the light quark (antiquark) in the meson, and the coefficients in the last two line are determined by the condition of the same normalization of the quark-antiquark states and of the orthogonality of the $(3/2)^+$ states to those in the $(1/2)^+$ doublet. 

Using the expressions in Eqs.(\ref{dec}) one can write formulas for the $J^{PC}=1^{--}$ $S$ wave states of the meson antimeson pairs in terms of four-quark wave functions, and after a Fierz transformation, relate these wave functions to the eigenstates $S_H^{PC}$ of the total spin $S_H$ of the heavy quark-antiquark pair and the eigenstates $J^{PC}_L$ of total angular momentum $J_L$ of the light degrees of freedom, with the latter including the spins of the light quark and antiquark, as well as any orbital motion of the (anti)quarks. For the $(1/2)^+ + (1/2)^-$ meson pair the result reads as
\bea
 {B^*_i \bar B_0 - \bar B^*_i B_0 \over \sqrt{2}}&:& \quad {1 \over 2} \, 1^{--}_H \otimes 0^{++}_L +{1 \over \sqrt{2}} \, 1^{--}_H \otimes 1^{++}_L + {1 \over 2} \, 0^{-+}_H \otimes 1^{+-}_L~, \nonumber \\
 {B_{1i} \bar B - \bar B_{1i} B \over \sqrt{2}} &:& \quad {1 \over 2} \, 1^{--}_H \otimes 0^{++}_L -{1 \over \sqrt{2}} \, 1^{--}_H \otimes 1^{++}_L + {1 \over 2} \, 0^{-+}_H \otimes 1^{+-}_L~, \nonumber \\
 {i \over 2} \, \epsilon_{ijk} \, \left ( B_{1j} \bar B^*_k - \bar B_{1j}  B^*_k \right ) &:& \quad - {1 \over \sqrt{2}} \, 1^{--}_H \otimes 0^{++}_L + {1 \over \sqrt{2}} \,  0^{-+}_H \otimes 1^{+-}_L~.
\label{one1}
\eea
For the  $(3/2)^+ + (1/2)^-$ pairs the corresponding formulas are the following
\bea
{B^*_{1i} \bar B - \bar B^*_{1i} B \over \sqrt{2}} &:& ~~ {i \over 2 \, \sqrt{2}} \, 1^{--}_H \otimes 1^{++}_L + {\sqrt{5} \over 2 \sqrt{2} } \, 1^{--}_H \otimes 2^{++}_L + {i \over 2} \, 0^{-+}_H \otimes 1^{+-}_L~, \nonumber \\
{i \over 2} \, \epsilon_{ijk} \, \left ( B^*_{1j} \bar B^*_k - \bar B^*_{1j}  B^*_k \right ) &:& ~~ {3 i \over 4} \, 1^{--}_H \otimes 1^{++}_L - {\sqrt{5} \over 4 } \, 1^{--}_H \otimes 2^{++}_L + {i \over 2 \sqrt{2}} \, 0^{-+}_H \otimes 1^{+-}_L~, \nonumber \\
 \sqrt{3 \over 10} \, \left ( B^*{2ij} \bar B^*j - \bar B^*{2ij}  B^*j \right ) &:& ~~  {i \, \sqrt{5}  \over 4} \, 1^{--}_H \otimes 1^{++}_L + {1 \over 4 } \, 1^{--}_H \otimes 2^{++}_L - {i \, \sqrt{5}\over 2 \sqrt{2}} \, 0^{-+}_H \otimes 1^{+-}_L . 
\label{one3}
\eea
It should be noted that the heavy-quark wave functions in Eqs.(\ref{one3}) are different from those in Eqs.(\ref{one1}), since the $(3/2)^+$ doublet of mesons is split from the $(1/2)^+$ by a gap that is parametrically of order $\Lambda_{QCD}$. For this reason the transformations described by Eqs.(\ref{one1}) and (\ref{one3}) involve separate unitary $3 \times 3$ matrices.

The electromagnetic current of the heavy quarks, $(b^\dagger \sigma_i b)$ produces only the $1^{--}_H \otimes 0^{++}_L$ state, since the light degrees of freedom are initially in the vacuum $0^{++}$ state. One can readily see that this component is absent in the decomposition (\ref{one3}) so that $S$ wave production of the $(3/2)^+ + (1/2)^-$ pairs is impossible without changing the spin state of the heavy quark-antiquark pair~\cite{lv}. Furthermore, the $1^{--}_H \otimes 0^{++}_L$ state enters with the same projection coefficient in the first two lines of Eqs.(\ref{one1}). Thus in the limit, where the pairs $B^*  B^0$ and $B B_1$ are degenerate in mass, only their coherent combination
\be
{1 \over 2} \left ( B^*_i \bar B_0 - \bar B^*_i B_0 + B_{1i} \bar B - \bar B_{1i} B \right ) :~~ {1 \over \sqrt{2}} \, 1^{--}_H \otimes 0^{++}_L + {1 \over \sqrt{2}} \, 0^{-+}_H \otimes 1^{+-}_L
\label{full11}
\ee
would be produced in $e^+e^-$ annihilation, while the orthogonal combination 
\be
{1 \over 2} \left ( B^*_i \bar B_0 - \bar B^*_i B_0 - B_{1i} \bar B + \bar B_{1i} B \right ) :~~  1^{--}_H \otimes 1^{++}_L
\label{full12}
\ee
decouples from the electromagnetic current of $b$ quarks. The very close degeneracy in the mass of the $(1/2)^+ + (1/2)^-$ strange mesons is known for the charmed mesons: the masses of the pairs $D_{s0}(2317) D^*_s$ and $D_{s1}(2460) D_s$ differ by only approximately $2 \pm 1\,$MeV around the mean value 4429\,MeV. As will be discussed further in this text, one can expect a similar degeneracy for the corresponding beauty-strange mesons.

The heavy-light spin structure in Eqs.(\ref{one1}) is written for the pairs of free non-interacting mesons. In the heavy quark limit the spin of the heavy quarks decouples and the interaction depends only on the angular momentum state $J^{PC}$ of the light components. As can be readily seen, the diagonal interaction in the states in the first two lines in Eqs.(\ref{one1}) is identical, so that the off-diagonal interaction results in that the eigenstates of the Hamiltonian are their sum and the difference, as indicated in Eqs.~(\ref{full11}) and (\ref{full12}). Thus the latter two states are the ones that are generally split by the interaction between the mesons.

It can be noted that the heavy-light spin structure of the states in Eq.(\ref{full11}) and that in the last line of Eq.(\ref{one1}) is quite similar to that~\cite{bgmmv} of the exotic bottomonium-like resonances $Z_b(10610)$ and $Z_b(10650)$ located respectively at the thresholds of $B^* \bar B$ and $B^* \bar B^*$ pairs of non-strange $B$ mesons, albeit with opposite parity and, naturally, with different light flavor properties. The presence of components with the total spin of the heavy quark pair with both $S_H=1$ and $S_H=0$ is used to explain comparable rates of decay of the $Z_b$ resonances to the final states with ortho- bottomonium: $Z_b \to \Upsilon(nS) \pi$ with $n=1,2,3$, and with para- bottomonium: $Z_b \to h_b(kP) \pi$ with $k=1,2$. Thus one can expect that a similar enhancement of (an apparent) HQSS breaking takes place near the threshold of $B_{s0}B^*_s$ ($B_{s1} B_s$) and near the higher threshold of $B_{s1} B^*_s$. This behavior can be observed e.g. through a production in $e^+e^-$ annihilation of e.g. final states $\chi_b(1P) \phi$ as well as $h_b(kP) \eta$, or $h_b(1P) \eta'$. There is however a potentially important difference of the discussed states with hidden-strangeness from the isovector $Z_b$ mesons. Namely, the former ones have no net light flavor and can thus mix with pure bottomonium $b \bar b$, while the $Z_b$ resonances can not. The $J^{PC}=1^{--}$ states of the bottomonium that can mix with the considered meson pairs have the spin structure $1^{--}_H \otimes 0^{++}_L$, and such mixing would modify, compared to Eq.(\ref{full11}) the relative weight of the spin-triplet and spin-singlet states of the $b \bar b$ pair. Thus  relative yield of production of ortho- and para-bottomonium with light mesons does depend on such possible mixing. However to an extent that the considered hidden-strangeness meson are present, the discussed effects of near-threshold HQSS breaking should be observable.

The presence of the excited heavy mesons can be significantly obscured if there is a strong rescattering of the considered two-meson states to final states with pairs of lower mass heavy mesons. In particular this strong mixing of channels may be the reason that a production in $e^+e^-$ annihilation of excited strange charmed mesons is not evident in the charmonium-like energy range of 4.4 - 4.6\,GeV. It can however be argued that the rescattering should be significantly weaker for the pairs of bottom mesons, at least for that into a pair of lower-lying mesons. Indeed, the threshold for a pair containing an excited meson is above that for lower mass mesons by an energy gap $\Delta$ which in the limit of heavy quark mass is a constant of order $\Lambda_{QCD}$. Thus the momentum of the lighter mesons near the threshold for a pair with an excited meson is $p \approx \sqrt{M \Delta}$, with $M$ being a heavy meson mass. Clearly, the momentum scales as $\sqrt{M_Q}$ and becomes large in the heavy quark limit, so that short-distance QCD counting, based on hard gluon exchange, can be applied. Thus the rescattering amplitude, e.g. of $(1/2)^+ + (1/2)^- \to (1/2)^- + (1/2)^-$ scales as at least the second inverse power of the momentum transfer,
\be
A \left [ \left ( {1 \over 2} \right )^+  + \left ( {1 \over 2} \right )^- \to \left ( {1 \over 2} \right )^- + \left ( {1 \over 2} \right )^- \right ] \propto p^{-2} \propto M_Q^{-1}~,
\label{sca}
\ee
so that for the $b$ hadrons such rescattering should be suppressed. 

The argument based on large momentum transfer generally does not apply to the scattering into final states with one or few extra pions, where the heavy mesons are slow and the energy excess is carried by the pions. However, in absolute terms the gap $\Delta$ is only at most 0.5 - 0.6 GeV, and it would be quite unnatural if multiparticle configurations  dominated the produced final states at such moderate energy release. Neither a dominance of such final states is observed in the corresponding charmonium-like energy range. 

Furthermore, the presented argument does not work, if there is a meson-antimeson channel  containing an excited  $B$ meson, whose threshold happens to be numerically close to that for the considered channels with hidden strangeness. This appears to be the case in the charmonium-like sector, where the mass of the pair $D^*_2(2460) \bar D+c.c.$ happens to be just about 100\,MeV below that for the excited hidden strangeness meson pair, which may be the reason for a substantial yield of this $D$ wave final state in vicinity of the $\psi(4415)$ peak. As discussed further a similar `coincidence' can also occur for the bottomonium-like states, so that the importance of the channel mixing still remains an open issue.

The significance of the coherence of the channels with beauty-strange excited mesons and of the heavy-light spin structure in Eq.(\ref{full11}) depends on the mass degeneracy of the meson pairs $B_{s0} B^*$ and $B_{s1} B$, implying essentially equal spin splitting of mass in the $(1/2)^+$ and $(1/2)^-$ doublets. This degeneracy is well known for the charmed-strange mesons: $M(D_{s1}) - M(D_{s0}) =  M(D_s^*) - M(D_s)+ O(1 \, {\rm MeV})$. An approximate equality of these mass splitting was argued~\cite{beh} from a chiral theory approach, where the $(1/2)^+$ and $(1/2)^-$ doublets are considered as parity partners. In particular, this approach implies that a similar, if not better degeneracy should hold for the beauty-strange mesons:
\be
 M(B_{s1}) - M(B_{s0}) =  M(B_s^*) - M(B_s) + O(1 \, {\rm MeV}) \approx 48\,MeV~.
\label{dm}
\ee

One can also present a purely phenomenological argument, although a somewhat circumstantial, for this mass relation to hold. Indeed, the mass splittings in a heavy-quark spin doublet is inversely proportional to the heavy quark mass $M_Q$ with possible higher-order terms  (see e.g. in the review \cite{bsu}). Thus the ratios of the splittings in the doublets of beauty and charmed mesons should all be the same, modulo possible higher-order corrections. If the latter corrections are indeed small, the validity of the relation (\ref{dm}) follows automatically from the known degeneracy of the charmed-strange mesons. The accuracy of a strict proportionality between the spin splittings of the $b$ and charmed hadrons can be tested using the recently measured masses of  excited $(3/2)^+$ bottom mesons~\cite{pdg}, both with and without strangeness: $M(B^{*0}_1) = 5726.0 \pm 1.3\,$MeV, $M(B^{*0}_2) = 5739.5 \pm 0.7\,$MeV,  $M(B^*_{s1}) = 5828.63 \pm 0.27\,$MeV and $M(B^*_{s2}) = 5839.84 \pm 0.18\,$MeV. Using exact proportionality of the splittings, the difference between these masses could be predicted from the known splitting in the charmed $(1/2)^-$ and $(3/2)^+$ doublets, and that for the $(1/2)^-$ bottom mesons. In the sector with strangeness the relation reads as
\be
M(B^*_{s2}) - M(B^*_{s1}) = { M(B_s^*) - M(B_s) \over M(D_s^*) - M(D_s) } \, \{ M[D^*_{s2}(2573)] - M[D^*_{s1}(2536)] \} = 11.5 \pm 0.5\,{\rm MeV}~,
\label{preds}
\ee
while for the non-strange mesons the relation is
\be
M(B^{*0}_2) - M(B^{*0}_1) = { M(B^{*0}) - M(B^0) \over M(D^{*0}) - M(D^0) } \, \{ M[D^*_2(2460)^0] - M[D^*_1(2420)^0] \} = 12.7 \pm 0.2\,{\rm MeV}~.
\label{pred}
\ee
The numerical values in these estimates are in full agreement, within the experimental uncertainty,  with the   measured mass splittings in the $(3/2)^+$ bottom meson doublets (respectively $11.2 \pm 0.3\,$MeV and $13.5 \pm 1.5\,$MeV). It is certainly (logically) possible that the contribution of unknown higher terms in $M_Q^{-1}$ is different in the $(1/2)^+$ doublet than in the $(3/2)^+$. However, such difference would present an intriguing problem of its own.

Prediction of the absolute position of the center-of-gravity (c.o.g.) of the beauty-strange $(1/2)^+$ doublet $\bar M[B_s(1/2)^+] = [3 \, M(B_{s1})+ M(B_{s0})]/4$ by extrapolation from the charmed sector involves an uncertainty resulting from $O(M_Q^{-1})$ contribution of the heavy quark kinetic energy to the meson mass~\cite{bsu}.  For the non-strange excited $B$ mesons the c.o.g. $\bar M[B(3/2)^+] = [3 \, M(B^*_{1})+ 5 M(B^*_{2})]/8=5734.4 \pm 0.7\,$MeV is above the c.o.g. for the ground-state $(1/2)^-$ mesons by $\Delta[B(3/2)^+]= 421.0 \pm 0.7\,$MeV, which is noticeably smaller than the similar difference for the $D$ mesons, $\Delta[D(3/2)^+]= 474.3 \pm 0.2\,$MeV. For the strange heavy mesons the c.o.g. of the $(3/2)^+$ doublet is at $\bar M[B_s(3/2)^+] = 5835.63 \pm 0.15\,$MeV, so that $\Delta[B_s(3/2)^+]= 436.4 \pm 1.4\,$MeV, while $\Delta[D_s(3/2)^+]= 480.3 \pm 0.6\,$MeV. In both cases the difference between the excitation energy in the charm and the bottom sector is in the same ballpark as the uncertainty of $\pm 35\,$MeV attributed in Ref.~\cite{beh} to the $O(M_Q^{-1})$ effects.

Although the discussed coherence of two channels in $e^+ e^-$ annihilation and appearance of HQSS breaking processes near the thresholds for $B_{s0} \bar B_s^*$ ($B_{s1} \bar B_s$) and $B_{s1} \bar B_s^*$ should be expected even if there is no threshold singularities for these states, the experimental visibility would be greatly enhanced if $S$ wave molecular peaks existed in these channels. Karliner and Rosner~\cite{kr} recently considered the forces due to $\eta$ exchange in exotic molecular states, and concluded that in the discussed here $C$-odd states with hidden strangeness this force corresponds to a repulsion. However the $\eta$ exchange is only one mechanism (not necessarily dominant) of interaction and shallow bound $S$ wave states may arise due to other forces~\footnote{The currently available data of a scan~\cite{babarscan} of the total cross section up to 11.2\,GeV total energy show no significant features above the $\Upsilon(6S)$ peak at approximately 11.0\,GeV. However, peaks may appear in specific exclusive channels, as is in the well known case of $Y(4260)$, where the total cros section has a local minimum, and still there is a peak in the yield of $J/\psi \pi \pi$.}. Also, as previously mentioned, properties of such states can be affected by mixing with $J^{PC}=1^{--}$ states of pure bottomonium. It is not clear at present to what extent the behavior observed at in the corresponding threshold region of charmonium-like states can serve as a guidance for expectations in the bottomonium-like sector. Indeed the hidden-charm peak $\psi(4415)$ (with the actual mass ~\cite{pdg} $4421 \pm 4\,$MeV) can be a highly excited $S$ wave level of pure bottomonium effectively masking a smaller contribution of the threshold states of $D_{s0} \bar D_s* + c.c.$ and $D_{s1} \bar D_s$. Furthermore, even if this peak contains a significant admixture of the discussed $S$ wave mesons pairs, the presence of hidden strangeness can still be obscured by strong coupling to channels with lighter non-strange charmed mesons. It can be noted, however, that a certain activity in the channels with charmed strange mesons has been observed~\cite{babar,belle} around the $\psi(4415)$ peak. Furthermore, recent data~\cite{besiii} suggest a possible enhancement of production of the HQSS breaking channel $h_c \pi \pi$ in the same energy range. However, the HQSS is not as strong in the charmonium-like sector as in the bottomonium-like due to smaller mass of the charmed quark, and the same final state is also observed at lower energies. In connection with the uncertainty of the status of hidden strangeness production in the $\psi(4415)$ energy region, new data at the threshold of $D_{s1} (2460) \bar D^*_s$ i.e. around 4.57\,GeV may prove to be helpful for exploring the yield of $(1/2)^+$ excited charmed-strange mesons in $e^+e^-$ annihilation.

In summary. It has been shown that in the $S$ wave production in $e^+e^-$ annihilation of the channels containing excited beauty-strange mesons only one coherent combination of the lower mass pairs, described by Eq.(\ref{full11}), is present due to HQSS. The heavy-light spin structure of this combination as well as of the heavier pair $D_{s1} \bar D^* - c.c.$, suggests that processes with (apparent) HQSS violation, such as e.g. the production of $h_b(kP) \pi \pi$, $h_b{kP} \eta$, $h_b(1P) \eta'$ should proceed near the corresponding two thresholds with rate comparable to those allowed by the HQSS. The presence of excited hidden beauty-strange mesons at these energies is likely to be more conspicuous than at the corresponding energy in the charmonium-like sector due to suppression of mixing of the produced states with two-body pair of the ground state mesons with or without hidden strangeness, described by Eq.(\ref{sca}). This suppression however does not apply to the final states with low-momentum heavy mesons, e.g. to the channels with excited non-strange $B$ mesons, such as $B^*_2(5740) \bar B + c.c.$. The significance of hidden-strangeness conversion into such channels can not be predicted theoretically at present, and may require an experimental study that will be feasible at Belle II.

I gratefully acknowledge illuminating discussions with Alexander Bondar.
This work is supported in part by U.S. Department of Energy Grant No.\ DE-SC0011842.

\end{document}